\newcommand{\beq}{\begin{equation}}
\newcommand{\enq}{\end{equation}}
\newcommand{\bee}{\begin{eqnarray}}
\newcommand{\ene}{\end{eqnarray}}
\newcommand{\bem}{\begin{mathletters}}
\newcommand{\enm}{\end{mathletters}}
\newcommand{\non}{\nonumber}

\documentstyle[aps,prl,twocolumn,epsfig]{revtex}

\begin{document}
\draft
\preprint{HEP/123-qed}
\twocolumn[\hsize\textwidth\columnwidth\hsize\csname
@twocolumnfalse\endcsname

\title{On quantization of weakly nonlinear lattices. Envelope solitons.}

\author{V. V. Konotop$\dag$\cite{email1} and  S. Takeno$\ddag$}

\address{
$\dag$ Departamento de F\'{\i}sica da Faculdade de Ci\^encias and
Centro de F\'{\i}sica da Mat\'eria Condensada, Universidade de
Lisboa, Complexo Interdisciplinar, Av. Professor Gama Pinto 2,
P-1649-003, Lisbon, Portugal.
\\
$\ddag$ Faculty of Information Science, Osaka Institute of
Technology,
1-79-1 Hirakata, Osaka 573-01, Japan
}
\date{\today}
\maketitle

\begin{abstract}
A way of quantizing weakly nonlinear lattices is proposed. It is
based on introducing "pseudo-field" operators. In the new
formalism quantum envelope solitons together with phonons are
regarded as elementary quasi-particles making up boson gas. In the
classical limit the excitations corresponding to frequencies above
linear cut-off frequency are reduced to conventional envelope
solitons. The approach allows one to identify the quantum soliton
which is localized in space and understand existence of a narrow
soliton frequency band.
\end{abstract}

]

One of the most important tasks of the modern physics of atomic
lattices is experimental observation of nonlinear localized
excitations which are believed to exist. Although in specially
prepared chains of coupled oscillators such excitations have been
obtained experimentally \cite{exp} in the microscopic world their
observation is hardly possible in a direct way. The first report
on observation of intrinsic localized modes, although in a
magnetic model, appeared only  recently \cite{sievers}. This rises
a problem of construction of statistical mechanics of nonlinear
lattices which coveres both linear and nonlinear excitations and
allows one to describe contribution of the localized modes to
macroscopic (measurable) characteristics of solids. As the first
step one has to provide quantization of atomic chains.

Anharmonic quantum lattices have been considered in literature.
Two approaches developed so far one can classify following
\cite{NumbState} as  the number state method \cite{NumbState} and
the quantum inverse scattering tecnique \cite{QISM}. Another
method based on exact numerical diaganolization of a hamiltonian
has been developed in \cite{WGB,WBGS} where breather modes were
obtained in a lattice with $\phi^4$ on-site potential  \cite{WGB}
and in a coupled electron-phonon system \cite{WBGS}, both are
nonintegrable. The breather modes are identified through the
spectrum and dynamic correlation functions.

The mentioned approaches deal with  nonlinear quantum hamiltonians
given {\it apriori}.  An alternative and somehow complementary way
of introducing quantum systems is quantization which starts with a
classical hamiltonian. As it is evident, the canonical
quantization, i.e. substitution of c-numbers by their operators
satisfying canonical commutation relations, formally can be
provided in the case of nonlinear lattices, as well. However, in a
generic situation this way is not tractable neither  analytically
nor numerically.

At the same time, methods of analytical description of nonlinear
classical lattices are well elaborated today. They are based on
introducing different small parameters and result in various
spatially localized excitations. Among them we mention envelope
solitons (ES) (see e. g. \cite{Kon}) which appear  when the wave
amplitude is considered as a small parameter and intrinsic
localized modes or breathers (see e. g. \cite{st}). In the case
where a small parameter can be identified as coupling between
neighbor sites a quantization can be performed in a way proposed
in \cite{AubQuant}. The role of the coupling constant magnitude
in the context of quantum lattices has been discussed in
\cite{WGB}.

In the present Letter we concentrate on quantization of a
nonlinear lattice when the small parameter is the nonlinearity,
and thus allowing in classical limit existence of ES. The
respective physical limit can be interpreted as opposite to one
considered in \cite{AubQuant}. Also in contrast to the previous
studies of nonintegrable quantum nonlinear lattices, ES appears
to be a dynamical object and thus allows one to construct a
kinetic theory of interacting solitons and phonons.

There are several points to be reflected by the theory. It has
been established that quantum nonlinear lattices have well
pronounced soliton band in the spectrum which is extremely narrow
for a given nonlinearity \cite{NumbState}. This means that the
respective object could be unambiguously identified either as ES
or as intrinsic localized mode depending on the value of the
frequency. Another point is the identification of a quantum ES
which must go beyond computing the frequency along. Indeed, an
important difference between ES and a phonon is that the soliton
is spatially localized while the phonon is localized only in the
momentum space. Thus we intend to obtain a quantum soliton as an
object, a quasi-particle, localized  in space (a possible way of
construction of such wave packets is described in
\cite{NumbState}). This is contrast to the quantization used the
field theory \cite{field} where a kink, being a topological
object, is considered as a vacuum state.  Finally, an important
property of the classical theory is that ES are governed in the
leading order by the nonlinear Schr\"{o}dinger equation \cite{Kon}
what means that in the leading order they do not interact with
linear phonons and can be regarded as independent quasi-particles.
Thus in the leading order the quantum theory must allow one to
introduce creation and annihilation operators for quantum soliton
and neglect its interactions with other quasi-particles. In this
way the interaction hamiltonian will naturally appear as a
perturbation and thus the gas of interacting quasi-particles --
solitons and phonons -- can be considered within the conventional
perturbation technique of the field theory.

It is to be mentioned that the approach developed below have some
points similar to one proposed in \cite{Haus} for quantum optical
solitons. In particular, quantum ES in \cite{Haus}is described in
terms of the field operators (we call them pseudo-field operators)
which are the Fourier transform of the conventional creation and
annihilation operators. There are however several essential
differences between our method and the approach of \cite{Haus}
(apart from different systems to be quantized). First, we start
with the original (i.e not approximated) classical dynamical
system. Second, we introduce a soliton as a quasi-particle. Third,
 we consider a highly rarified gas of
quasi-particles in the quasi-classical limit (while in \cite{Haus}
the Hartree approximation for a photon gas was exploited).
Finally, quantum ES emerges as a matrix element of the creation
operator applied to the vacuum state \cite{Wadati} (notice that in
the classical theory envelope solitons are not necessarily real).
If however one considers an ES as an excitation of the
displacement field (see $u_n$ below) then the quantum ES appears
to be an observable.

Speaking about quantization of ES one can identify one essential
feature of the quantum model. In the classical theory the ES
amplitude (and frequency detuning) is not a fixed value (see e.g.
\cite{Kon}). That is why, the energy associated with the classical
ES is not a fixed quantity. The respective quantum problem has a
scale defined by the Planck constant: it is natural to expect that
the excitation characterized by the frequency $\omega$ will have
the energy $\hbar\omega$. Taking into account that the linear
oscillator energy is given by $m\omega^2 A^2$ (here $m$ is a mass
and $A$ is the amplitude) one concludes that the amplitude of the
quantum ES
 must be of order of $\lambda=\sqrt{\hbar/\omega m}$. On the other hand
the amplitude of the soliton must be small enough. In order to
specify this last requirement, let us assume that the interaction
energy of neighbor sites is $U(x)$ and define a characteristic
scale of the energy variation: $L\sim
(|\frac{1}{U(x)}\frac{d^2U(x)}{dx^2}|)^{-1/2}$. Then the relation
between the nonlinear and linear terms in the classical model is
of order of $(A/L)^2$. It is this value that appears to be a small
parameter in the classical theory, or more precisely $A/L\ll 1$.
Then substituting the estimate for the amplitude obtained above on
the basis of quantum approach one gets the requirement
$\frac{\hbar}{\omega m L^2}\ll 1$ as a condition for the validity
of the small amplitude expansion. Scaling out all the quantities
and assuming that the characteristic spatial size of the
excitation is one, $L\sim 1$ we conclude that the effective small
parameter of the problem is $\hbar$, i.e. we a re dealing with the
quasiclassical limit.

We consider quantization of a one-dimensional monoatomic lattice
described by the hamiltonian
\begin{eqnarray}
\label{a1}
H=\frac 12 \sum_n\frac{p_n^2}{m}+
\frac 12 \sum_{n_1,n_2}K_2(n_1,n_2) x_{n_1} x_{n_2}
\nonumber \\
+
\frac 14 \sum_{n_1,...,n_4}K_4(n_1,...,n_4) x_{n_1}... x_{n_4}
\end{eqnarray}
where $x_n$ and $p_n=m\dot{x}_n$ are displacements of atoms from
the equilibrium position and their linear momenta, $K_2(n_1,n_2)$
and $K_4(n_1,...,n_4)$ are linear and nonlinear real force
constants. Overdot hereafter stands for the derivative with
respect to time. We concentrate on the case of the force
coefficients symmetric with respect to permutation of the indexes,
e.g. $K_2(n_1,n_2)=K_2(n_2,n_1)$. Also it is assumed that
$
\sum_{n_1}K_2(n_1,n_2)=0.
$

First of all we recall some facts of the theory of linear
lattices.  Consider vibrations of a one-dimensional (possibly disordered)
lattice described by the Hamiltonian
\begin{equation}
\label{a1*}
H=\frac 12 \sum_n\frac{p_n^2}{m_n}+\frac 12 \sum_{n,l}K(n,l) x_{n}
x_{l}
\end{equation}
where $m_n$ is the mass of the $n$th atom and
$K(n,l)=K(l,n)$ are real force constants.
 It is convenient to introduce
new dependent variables $ u_n=\sqrt{m_n}x_n $ and force constants
$ J(l,n)=K(l,n)/ \sqrt{m_lm_n}$. The respective   equation of
motion reads
\begin{equation}
\label{e2} \ddot{u}_n+\sum_{l}J(n,l)u_l=0
\end{equation}

It will be assumed that the lattice consists of ${\cal N}$, ${\cal
N}\gg 1$, atoms  and is subject to the cyclic boundary conditions.
Then equation (\ref{e2}) can be associated with the linear
spectral problem [$\psi_q(n)=\psi_q(n+{\cal N})$]
\begin{equation}
\label{e3}
\sum_{l}J(n,l)\psi_q (l)=\omega_q^2\psi_q(n),
\end{equation}
where real eigenvalues $\omega_q^2$ are squared eigenfrequencies
and $q$ denotes eigenmodes, eigenfunctions $\psi_q(n)$ and
$\bar{\psi}_q(n)$ correspond to the same eigenfrequency (an
overbar stands for complex conjugation). Introducing the matrix
${\bf J}$ with the element $J(n,l)$ placed at $n$th row and $l$th
column,   a ket-vector $|q\}=\mbox{col}(...,\psi_q(n-1),
\psi_q(n), \psi_q(n+1),...)$
 and a bra-vector
$\{ q|=(...,\bar{\psi}_q(n-1), \bar{\psi}_q(n), \bar{\psi}_q(n+1),...)$
 we rewrite the eigenvalue problem
(\ref{e3}) in the form $ {\bf J}|q\}=\omega_q^2|q\}$.
We assume that $\omega_q=\omega_{-q}$, i. e.
$ \bar{\psi}_q(n)=\psi_{-q}(n)$. $\psi_q(n)$ constitute an
orthonormal complete set
$
\{
q|q'\}=\sum_{n}\bar{\psi}_q(n)\psi_{q'}(n)
=\delta_{q,q'}
$, and $
\sum_q
\bar{\psi}_q(n)\psi_q(l)=\delta_{n,l}
$ (here $\delta_{l,n}$ is the Kronecker delta).

Quantization of lattice (\ref{e2}) can be reduced, first, to the
problem of the diagonalization of the matrix ${\bf J}$ and,
second, to the quantization of independent linear oscillators. In
the normalized variables the kinetic energy is
 written in the form $\{\pi|{\bf I}|\pi\}$ where
$|\pi\}=$col$(...,\pi_{n-1},\pi_{n},\pi_{n+1},...)$,
$\pi_n=\dot{u}_n$ is the momentum conjugate to $u_{n}$, and ${\bf
I}$ is the unit matrix. Thus any matrix diagonalizing the
potential energy will preserve the diagonal form of the kinetic
energy. To provide diagonalization of the potential energy we
construct a ${\cal N}\times {\cal N}$ matrix ${\bf \Psi}=
(...,|q_{j-1}\}, |q_j\}, |q_{j+1}\},...)$ [here numbering of the
discrete eigenvalues is introduced]. As is clear ${\bf J}{\bf
\Psi}={\bf \Psi}{\bf \Omega}^2$, where ${\bf
\Omega}=$diag$(...,\omega_{q_{j-1}},\omega_{q_{j}},\omega_{q_{j+1}},...)$,
i.e. the matrix ${\bf \Psi}$ diagonalizes ${\bf J}$: ${\bf
\Psi}^{-1}{\bf J}{\bf \Psi}={\bf \Omega}^2$.  If one represents $
|u\}={\bf \Psi}|X\}$ and $\{ u|=\{ X|{\bf \Psi}^{-1}$ where
$|u\}=$col$(...,u_{n-1}, u_n, u_{n+1},...)$ and
$|X\}=$col$(...,X_{q_{j-1}}, X_{q_j}, X_{q_{j+1}},...)$
 is a column matrix of the normal coordinates
with $X_q=\bar{X}_{-q}$, the potential energy will take the form
$\{ X|{\bf \Omega}^2|X\}$. Finally
\begin{equation}
\label{a11} H=\frac 12 \{ P|P \} +\frac 12 \{ X|{\bf \Omega}^2 |X
\}\,,
\end{equation}
where
$|P\}=$col$(...,P_{q_{j-1}},P_{q_j},P_{q_{j+1}},...)$ is related to
the representation $|\pi\}={\bf \Psi}|P\}$.

In order to quantize the linear disordered lattice one introduces
phonon annihilation and creation operators
\begin{eqnarray}
\label{a12} \hat{a}_q=\sqrt{\frac{\omega_q}{2\hbar}}\hat{X}_q+
\frac{i}{\sqrt{2\hbar\omega_q}}\hat{P}_{-q} \nonumber \\
\hat{a}_q^{\dag}=\sqrt{\frac{\omega_q}{2\hbar}}\hat{X}_q-
\frac{i}{\sqrt{2\hbar\omega_q}}\hat{P}_q
\end{eqnarray}
where $\hat{X}_q$ and $\hat{P}_q$ are operators of the c-numbers
$X_q$ and $P_q$ satisfying the canonical commutation relations.
Then the Hamiltonian is reduced to the conventional form
$
\hat{H}_l=\sum_q \hbar\omega_q\left(\hat{a}_q^{\dag}
\hat{a}_q+\frac 12\right).
$
Operators of displacements and of linear momenta read \bee
\label{displ}
\hat{u}_n=\sum_q\sqrt{\frac{\hbar}{2\omega_q}}[\psi_q(n)\hat{a}_q+
\bar{\psi}_q(n)\hat{a}_q^{\dag}]\nonumber \\
\hat{\pi}_n=-i\sum_q\sqrt{\frac{\hbar\omega_q}{2}}[\psi_q(n)\hat{a}_q-
\bar{\psi}_q(n)\hat{a}_q^{\dag}] \ene

Let us introduce Schr\"{o}dinger {\em pseudo-field operators}
according to the relations \beq \label{psi}
\hat{\psi}_n=\sum_q\psi_q(n)\hat{a}_q,\,\,\,\,\,\,\,\,\,
\hat{\psi}^{\dag}_n=\sum_q\bar{\psi}_q(n) \hat{a}^{\dag}_q \enq
They describe creation and annihilation of a quasi-particle at the
site $n$. We emphasize the main distinguishing feature of these
operators compared with the canonical field operators: $\psi_q(n)$
is an eigenfunction of the classical problem rather than a
wave-function.

It is not difficult to verify that
\beq%
 \hat{a}_q=\sum_n\bar{\psi}_q(n)\hat{\psi}_n,\,\,\,\,\,\,\,\,
\hat{a}_q^{\dag}=\sum_n\psi_q(n)\hat{\psi}^{\dag}_n
\enq%
 and there
exist commutation relations
\bee%
\nonumber
 [\hat{\psi}_{n_1},\hat{\psi}_{n_2}]=
[\hat{\psi}^{\dag}_{n_1},\hat{\psi}^{\dag}_{n_2}]=0,
\qquad
[\hat{\psi}_{n_1},\hat{\psi}^{\dag}_{n_2}]=\delta_{n_1,n_2}
\ene%
The operator of number of quasi-particles is given by \beq
\label{N}
\hat{N}=\sum_q\hat{a}_q^{\dag}\hat{a}_q=\sum_n\hat{\psi}^{\dag}_{n}
\hat{\psi}_{n} \enq

Finally, following the conventional procedure we introduce
Heisenberg pseudo-field operator
\beq
\label{h_psi}
\hat{\Psi}_n(t)=
\exp\left(i\frac{\hat{H}t}{\hbar}\right)\hat{\psi}_{n}
\exp\left(-i\frac{\hat{H}t}{\hbar}\right)
\enq
It solves the equation
\beq
\label{evol}
\frac{\partial^2}{\partial
t^2}\hat{\Psi}_n(t)+\sum_l J(n,l)\hat{\Psi}_l(t)=0.
\enq

Let us now consider a nonlinear lattice with classical Hamiltonian
(\ref{a1}) after the renormalization $J_2(n_1,n_2)=K_2(n_1,n_2)/m$
and $J_{4}(n_1,...,n_4)=K_4(n_1,...,n_4)/m^2$. To obtain explicit
form of the quantum hamiltonian we assume that the definition
(\ref{displ}) holds in the nonlinear case. We substitute operators
of the displacement and linear momenta (\ref{displ}) into the
Hamiltonian, and express the result through the pseudo-field
operators, rearranging the last ones in the normal order and
dropping the constant which corresponds to the energy of the
lattice vacuum. Then $\hat{H}=\hat{H}_0+\hat{H}_{int}$. The
operator $\hat{H}_0$ has the form \bee \label{hamlin}
\hat{H}_0=\sum_{n_1,n_2}S_{n_1,n_2} \hat{\Psi}^{\dag}_{n_1}
\hat{\Psi}_{n_2}=\sum_q\hbar\omega_q\hat{a}_q^{\dag}\hat{a}_q \ene
with the kernel $S_{n_1,n_2}
=\sum_q\hbar\omega_q\psi_q(n_1)\bar{\psi}_q(n_2)$. Now $\psi_q(n)$
is an eigenfunction and $\omega_q$ is a frequency of the
"nonlinear" eigenvalue problem. The "nonlinearity contribution"
 emerges from the ordering procedure. More precisely,
the eigenvalue problem (\ref{e3}) is now considered with
the kernel
\beq
\label{J}
J(n,l)=J_2(n,l)+J_{d}(n,l)
\enq
where $J_2(n,l)$ is the force constant of the underline linear lattice
and the "deformation" $J_{d}(n,l)$ is given by
\[
J_{d}(n_1,n_2)=\frac{3\hbar}{2\omega_q}\sum_q\sum_{l_1,l_2}
J_4(n_1,n_2,l_1,l_2)\psi_q(l_1)\bar{\psi}_q(l_2).
\]

The interaction potential $\hat{H}_{int}$ has the form
\bee
\hat{H}_{int}=\sum_{n}S_{n_1n_2n_3n_4}[
\hat{\Psi}_{n_1} \hat{\Psi}_{n_2} \hat{\Psi}_{n_3} \hat{\Psi}_{n_4}
+\hat{\Psi}^{\dag}_{n_1} \hat{\Psi}_{n_2} \hat{\Psi}_{n_3}
\hat{\Psi}_{n_4}
\non \\ \non
+\hat{\Psi}^{\dag}_{n_1} \hat{\Psi}^{\dag}_{n_2} \hat{\Psi}_{n_3}
\hat{\Psi}_{n_4}+
\hat{\Psi}^{\dag}_{n_1} \hat{\Psi}^{\dag}_{n_2} \hat{\Psi}^{\dag}_{n_3}
\hat{\Psi}_{n_4}
+
\hat{\Psi}^{\dag}_{n_1} \hat{\Psi}^{\dag}_{n_2} \hat{\Psi}^{\dag}_{n_3}
\hat{\Psi}^{\dag}_{n_4}]
\ene
with the kernel
\bee
S_{n_1n_2n_3n_4}=
\frac{\hbar^2}{16}\sum_{q}\sum_{l}
\frac{J_4(l_1,l_2,l_3,l_4)}{\sqrt{\omega_{q_1}\omega_{q_2}
\omega_{q_3}\omega_{q_4}}}
\psi_{q_1}(l_1)\psi_{q_2}(l_2)
\times \non \\
\psi_{q_3}(l_3)\psi_{q_4}(l_4)
\bar{\psi}_{q_1}(n_1)
\bar{\psi}_{q_2}(n_2)\bar{\psi}_{q_3}(n_3)\bar{\psi}_{q_4}(n_4)
\non
\ene

In the small amplitude limit which corresponds to small $J_4$,
  $\hat{H}_{int}$ can be considered
as a perturbation.  That is why we introduce pseudo-field
operators in the interaction representation,
$\hat{\Psi}^{(0)}_n(t)$, by formula (\ref{h_psi}) where
$\hat{H}_0$ is used instead of $\hat{H}$. Then one can verify that
$\hat{\Psi}^{(0)}_n(t)$ solves (\ref{evol}) with $J(n,l)$ given by
(\ref{J}).

  In order to show how  ES comes out from the
above approach we notice that the ES solution corresponds to the
situation when only one quasi-particle is excited in the lattice.
In what follows quasi-particles corresponding to the frequencies
inside the spectrum band of the linear lattice,
$\omega\in[0,\omega_0)$ where $\omega_0$ is the cut-off frequency
of the underline linear lattice, are called phonons while
excitations with $\omega>\omega_0$ will be referred to as
solitons. Then
$
\hat{\Psi}=\hat{\Psi}_{ph}+\hat{\Psi}_{s}.
$

Let us introduce the notation $|n_{q_1},...,n_{q_N}\rangle$ for
the wave function of a state when $n_{q_j}$ quasi-particles with
the wave number $q_j$ are excited. Then the "one soliton"
 state, when there exists only one
eigenvalue $\omega_{s}$ bigger than $\omega_0$, is
$|0,...,0;1\rangle$  and
$
\psi_{s}(n)=\langle 0,...,0;0|\hat{\Psi}|0,...,0;1\rangle
$
[we use the notation $\psi_{s}(n)\equiv \psi_{q=\pi/a}$,
$\omega_{s}=\omega_{q=\pi/a}$]. In order to obtain $\psi_{s}$ we
rewrite Eq. (\ref{e3}) in the form
\bee \label{sol2} \omega_{s}^2\psi_{s}(n)=
\sum_m\psi_{s}(m)J_2(n,m) +\frac{3\hbar}{2\omega_{s}}\times \non
\\ \sum_{m_1,m_2,m_3} J_4(n,m_1,m_2,m_3)\psi_{s}(m_1)
\bar{\psi}_{s}(m_2)\psi_{s}(m_3)
\ene
As far as  the small amplitude limit is under consideration the
frequency $\omega_{s}$ is close to the cut-off frequency of the
underline linear lattice, $\omega_0$, and one can introduce a
small parameter $\epsilon$ ($\epsilon\ll 1$) through the relation
$
\omega_{s}^2-\omega_0^2=\epsilon^2\omega_0^2
$
where $\epsilon^2=2(\omega_{s}/\omega_0-1)$ is the frequency
detuning towards the forbidden zone.

We look for the solution of (\ref{sol2}) in the form
\beq \label{sol3}
\psi_{s}(n)=\epsilon\sqrt{\frac{2\omega_{s}}{3\hbar}}\varphi_n
A(\epsilon n) \enq
where $\varphi_n$ is the eigenvalue of the problem for the perfect
linear lattice, i.e. $J(n,l)=J_2(n,l)$ in (\ref{e3}). Then
expanding $A(\epsilon(n-l))$ (here $l$ depends on the number of
nearest neighbors interacting with given atom) in the Taylor
series with respect to $\epsilon l$, introducing $x=\epsilon an$
(treated as a continuum spatial variable) and taking into account
that $\psi_{s}(n-l)=\psi_{s}(n+l)$ one arrives at
\bee -\frac{a^2}{2}\sum_{n_1,n_2} (n_1-n_2)^2
J_2(n_1,n_2)\bar{\varphi}_{n_1}\varphi_{n_2}\frac{\partial^2
A}{\partial x^2}+ \non \\ \sum_{n}J_4(n_1,...,n_4) \varphi_{n_1}
\bar{\varphi}_{n_2} \varphi_{n_3} \bar{\varphi}_{n_4}|A|^2A \non
=\omega_0^2 A \ene
which is obtained for monochromatic solution of the 1D nonlinear
monoatomic lattice within the framework of the conventional
multiscale analysis \cite{Kon}.

The obtained results corraborate with the known ones. In
particular, it has been shown that for the given nonlinearity the
frequency of the quantum soliton is fixed by the normalization
condition \cite{Haus}. This condition defines $\epsilon$
introduced in (\ref{sol3}) and must be computed for each lattice
separately. It is however clear that in any case the result gives
$\epsilon\sim\sqrt{\hbar}$ which means the consistency of the
expansion. On the other hand this result does not contradict to
the existence of a narrow soliton band which has been explained in
\cite{NumbState,AubQuant,Haus} with help of different arguments.
Indeed, quantum ES is characterized not only by the "carrier wave"
frequency but also by the frequency smearing which is of order of
$(\hbar/m)^{3/2}\omega_0^{-1/2} L^3$. Our results corraborate also
with recent findings in the classical statistical mechanics of
nonlinear systems. Namely it is well known that boson display a
tendency of creating clusters (see e.g. \cite{Schwabl}). This
means that creation of a phonon and a quantum ES sufficiently
close to each other will end up in clustering of the two
quasi-particles. This can be viewed as absorption of the phonon by
the soliton. This phenomenon has been observed recently in
numerical experiments \cite{stat}. Finally, in the leading order
with respect to the nonlinearity solitons and phonons are
noninteracting objects. Interaction is taken into account in
higher orders of the expansion.

VVK is grateful to Prof. M. V. Berry for valuable comments and to
Prof. V. Eleonskii for attention to this work. We thank referee
for drawing our attention to Ref. \cite{Haus}. VVK acknowledges
support from FEDER and Program PRAXIS XXI, No
Praxis/P/Fis/10279/1998.

\end{document}